\documentclass[aps,twocolumn,superscriptaddress,showpacs,showkeys]{revtex4-1}
\usepackage{graphics,graphicx,dcolumn,bm,fleqn,epic,eepic,float,epsfig}
\usepackage{multirow,rotate,color,float}
\usepackage{amssymb,amsfonts,amsmath}
\usepackage{lipsum}
\usepackage{epstopdf}
\usepackage{times,natbib}
\usepackage{color}
\usepackage{soul}                    % package needed for overstriking
\definecolor{red}{rgb}{1,0,0}
\definecolor{green}{rgb}{0,1,0}
\definecolor{blue}{rgb}{0,0,1}

\begin{document}
\title{Rogue waves and entropy consumption} % For titles, only capitalize the first letter

\author{Ali Hadjihoseini}
\affiliation{Institute of Physics and ForWind, Carl-von-Ossietzky University of Oldenburg, Oldenburg, Germany}
\author{Pedro G.~Lind}
\affiliation{Institute of Physics, University of Osnabr\"uck, Osnabr\"uck, Germany}
\author{Nobuhito Mori}
\affiliation{Disaster Prevention Research Institute (DPRI), Kyoto University, Japan}
\author{Norbert P.~Hoffmann}
\affiliation{Hamburg University of Technology, 21073 Hamburg, Germany}
\affiliation{Imperial College, London SW7 2AZ, United Kingdom}
\author{Joachim Peinke}
\affiliation{Institute of Physics and ForWind, Carl-von-Ossietzky University of Oldenburg, Oldenburg, Germany}

\date{\today}

\begin{abstract}
  Based on data from the Japan Sea and the North Sea the occurrence of rogue waves is analyzed  by a scale dependent stochastic approach, which interlinks fluctuations of waves for different spacings. With this approach we are able to determine a stochastic cascade process, which provides information of the general multipoint statistics.  Furthermore the evolution of single trajectories in scale, which characterize wave height fluctuations in the surroundings of a chosen location, can be determined. The explicit knowledge of the stochastic process enables to assign entropy values to all wave events. We show that for these entropies the integral fluctuation theorem, a basic law of non-equilibrium thermodynamics, is valid. This implies that  positive and negative entropy events must occur. Extreme events like rogue waves are characterized as negative entropy events. The statistics of these entropy fluctuations changes with the wave state, thus for the Japan Sea %data
  the statistics of the entropies has a more pronounced tail for negative entropy values, indicating a higher probability of rogue waves.
  %compared to the wave state of the North Sea.
%We provide a thermostatistical description for explaining the emergence
%of rogue waves in the ocean. 
%In particularly, we show that the occurrence of rogue waves can be
%predicted from the distribution of entropy variations.
%Our analysis is supported by the integral fluctuation theorem, which
%we show to hold for data sets of sea-level fluctuations,
%and provides a framework for computing entropy from single scale trajectories.
%Scale trajectories of sea-level heights characterise wave height fluctuations
%and are introduced as a fundamental concept for explaining the emergence
%of rogue waves.
%Our method enables to properly predict if a rogue wave is likely to
%occur or not at a particular spot of the ocean.
%Finally, we also put our findings in perspective, towards a quantitative
%connection between the statistical description of a system out of equilibrium
%and its deterministic non-linear behaviour, bridging the gap between
%statistical approaches and coherent structures.
\end{abstract}



\maketitle

%\section{Introduction}

Oceanic rogue waves are usually defined as
extremely large waves that occur suddenly and
unexpectedly, even in situations where the ocean appears relatively
calm and quiet \cite{birkholz2015}.
While there are numerous reports from sailors claiming to have 
observed a rogue wave in the open ocean, 
rogue waves are very rare, which makes researching or forecasting
difficult \cite{Soto-Crespo2016}. %,Toenger2015}.
%Though there is no unique mathematical definition,
%a rogue wave can be defined as \cite{birkholz2015}
%% \cite{nature1,nature2,birkholz2015}
%a large amplitude wave that appears randomly and very rarely. 
Because of their size rogue waves can
be extremely dangerous, even to the large ocean liners,
appearing in different
forms of rare large amplitude events \cite{Onorato2013}.
% \cite{Efim2009,Onorato2013},
%as shown in Fig.~\ref{fig01}.
% \cite{Grimshaw2013,Alam2014,Toenger2015,Tong2016,Soto-Crespo2016}.
As a prototypical example of extreme events emerging in 
a stochastic ``background'', rogue waves have been 
investigated from various perspectives,
e.g.~using tools from non-linear waves and soliton
theory \cite{amin2011}. %,SciRep}.
% \cite{soliton,amin2011,SciRep}.
%%%%%%%%%%%%%%%%%%%%%%%%%%%%%%%%%%%%%%
\begin{figure}[b]
\begin{center}
\includegraphics[width=0.45\textwidth]{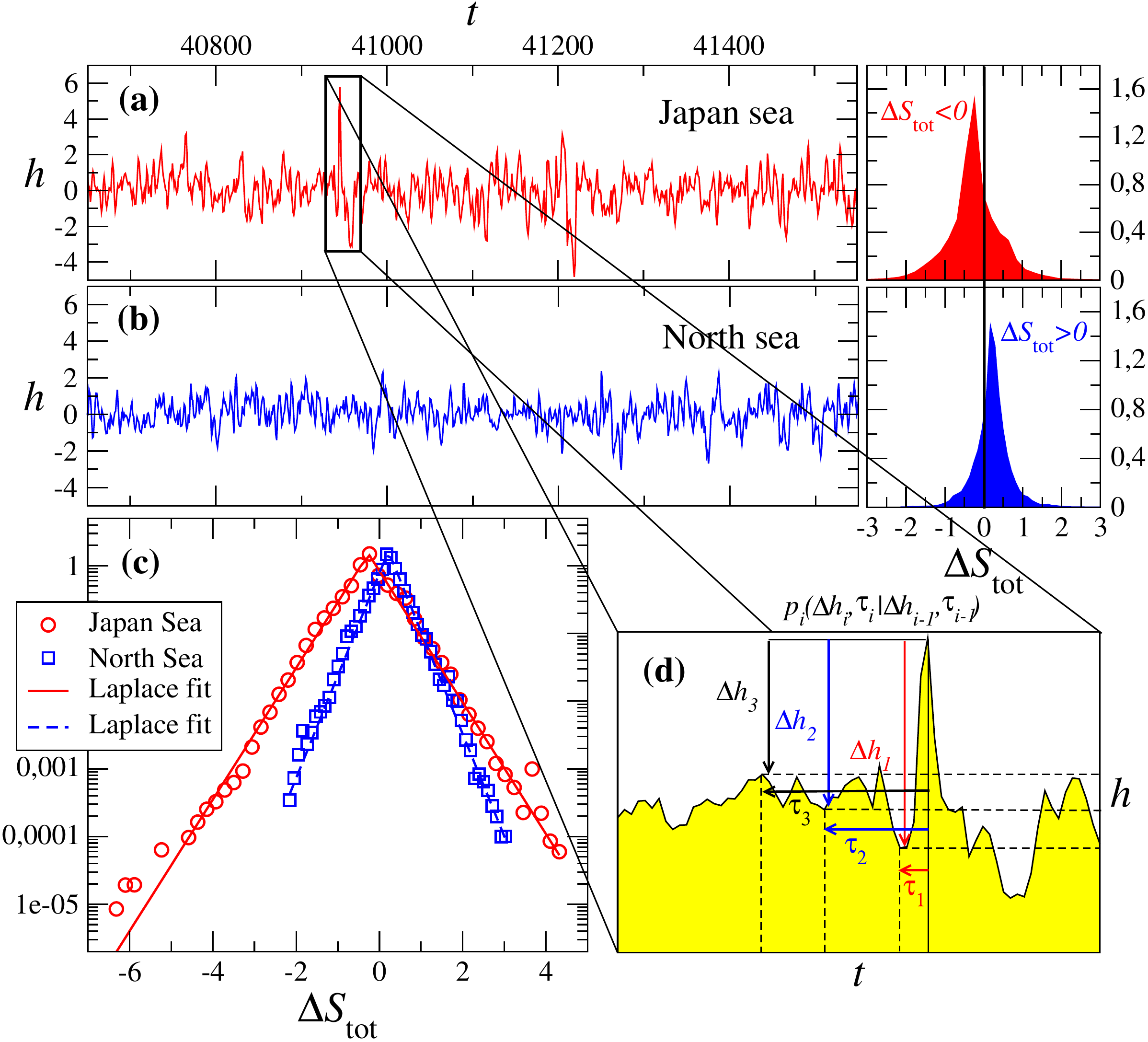}
\caption{\protect
  {\bf (a)} Series of sea surface height taken
  at the Japan Sea, where rogue waves are observed, 
  and {\bf (b)} surface heights from the North Sea,
  where rogue waves are not observed. 
  All heights are given in units of meter.
  On the right, one plots the distribution of entropies
  variations through trajectories associated to each series,
  which are well approximated by
  {\bf (c)} the Laplace distribution,
  with negative median for Japan sea.
  {\bf (d)} Illustration of the sea-level heights, as a
  scale process (see text).}
\label{fig01}
\end{center}
\end{figure}
%%%%%%%%%%%%%%%%%%%%%%%%%%%%%%%%%%%%%%

Due to the scarcity of observational data, many fundamental
questions are still under debate.
What exactly causes a specific rogue wave? Are there any fundamental
features of the ambient sea state
that lead to the occurrence of a rogue wave in the ocean? 
Is it possible to provide quantitative insight into how probable it is
to observe a rogue wave?
Often investigations into rogue waves are based
on models for wave packet evolution in non-linear
dispersive media \cite{Onorato2013}.
% \cite{nature1,nature2,Efim2009,Onorato2013,Osborne2000}.
Studies have been successful in demonstrating
the existence of rogue waves and also allowed classifying them into
different classes.
Still, the approach is fundamentally deterministic, while, as
the definition of rogue waves itself
suggests, a probabilistic description seems more natural.

In this paper we provide what is, to our knowledge,
the first evidence for thermodynamical processes
underlying the occurrence of rogue waves in nature.
Our findings do not contradict the findings from previous deterministic
approaches to investigate rogue waves, but instead complement the
present understanding with a thermostatistical perspective.
%following previous related 
%work HEREHERE \cite{nickelsen2013,seifert2005,Seifert2012}.
%,Jarzynski2017}. %,Jarzynski2011,

The findings reported here are based on the combination of two
important features that rogue waves share with extreme events
in general \cite{Hadjihosseini2014,Ali2016}:
first, they occur within short time-scales; second,
their large amplitude variations reflect a flow of energy
coming from the largest scales.
This energy flow through scales underlying the occurrence
of a rogue wave is similar to the picture of the energy flux in
Kolmogorov’s turbulence cascade \cite{kolmogorov}.
%Consequently,
Our work presents first evidence of a physical
connection between the emergence
of extreme events in systems far from equilibrium and the fundamental
features of energy flow in them.

%%%%%%%%%%%%%%%%%%%%%
%\section{From time processes to scale processes}

Two observational data sets are analyzed, one collected at
the Japan Sea, where rogue waves are observed,
and another collected at the North sea, where rogue waves
are almost absent \cite{Mori2002,Mori2002a}. %,FINOdata}.
Figures \ref{fig01}a and \ref{fig01}b sketch a part of these
two observational series.
The series from the Japan sea (Fig.~\ref{fig01}a)
includes the measured signal of one rogue wave with a height
of about $6$ meters.

We will show that the structure of rogue waves results from
an exchange of entropy between the wave environment and the
local wave condition itself, along a trajectory ``downscale''.
Furthermore, the distribution of the total entropy variation
along these downscale trajectories differ for the two cases:
in the data set where rogue waves are absent it has a positive
mode, whereas for the Japanese data set, which includes rogue
waves, the distribution mode is negative.
See the right plots of Figs.~\ref{fig01}a and \ref{fig01}b and
Fig.~\ref{fig01}c.
%Indeed, one main finding from our analysis is that
%rogue waves are characterised by a negative entropy production
%through a hierarchy of time scales. 

For the proper analysis of the stochastic series in
Fig.~\ref{fig01}, one aims at the derivation of
a predictor $h_{t+\tau}$ for the next time step $t+\tau$,
based on past measurements of the series, 
$\{ h_t, h_{t-\tau},\dots,h_{t-N\tau} \}$.
Here, $\tau$ is taken as the unitary time-lag between
successive measurements.
If the process is Markovian throughout its time evolution,
the predictor is a function of
the present state $h_t$ only, and the time series can be
statistically reproduced using the $2$-point statistics
$p(h_{t+\tau},h_t)$.
The propagator is then simply the conditional probability
density function 
$p(h_{t+\tau}\vert h_t)=p(h_{t+\tau},h_t)/p(h_t)$
with $p(h_t)=\int p(h_{t+\tau},h_t)dh_{t+\tau}$.
When the process is not Markovian, 
each value in the series depends on a larger set of previous values
and consequently one needs to extract a $N$-point statistics,
for larger $N>2$, which, in practice, is very cumbersome and challenging.
%in a straightforward way \cite{Friedrich1997}.
Ocean surface level time series turn
out not to be Markovian.

It is possible to overcome this shortcoming if one considers the 
concept of ``scale process''\cite{nickelsen2013,seifert2005,Seifert2012},
illustrated in Fig.~\ref{fig01}d.
A scale process describes how the variables' {\it increments}
$\Delta h_{k,t}:=h_{t}-h_{t-\tau_k}$
 change with $\tau_k:=k\tau$
for a fixed time $t$ and a chosen value of $\tau$. Here we denote time-lags as time-scales.
With such a concept, we now define the ensemble $\Delta h_{k,t}$
for fixed $t$ as a scale trajectory through time-scales
$\tau_k$ ($k= 1,\dots,N-1$).

To convert the series of sea-level height measurements in time
to its scale or increment counterparts has profound consequences, as
the latter now turns out to be  Markovian \cite{Hadjihosseini2014,Ali2016}.
As a consequence the full analysis of the data, based on the computation
of the $N$-point propagator $p(h_t\vert h_{t-\tau_1}, ... , h_{t-\tau_{N-1}})$ 
that predicts the time series of heights, can be decomposed into
$N-2$ increment propagators $p(\Delta h_{k,t}\vert \Delta h_{k+1,t},h_t)$
for each scale $k$ together with
the initial distribution, $p(\Delta h_0,\tau,h_t)$ \cite{Ali2016},
see also supplementary material I.

Each increment propagator can be extracted separately from
the time series \cite{Friedrich1997,Friedrich2011}, defining
a Fokker-Planck equation\cite{Risken} for the respective time-scale,
\begin{equation}
\begin{aligned}
-\tau\frac{\partial}{\partial\tau} p(\Delta h_k| \Delta h_{k'},h) =
\\ -\frac{\partial}{\partial \Delta h_k}\left[D^{(1)} (\Delta h_k,h) p(\Delta h_k| \Delta h_{k'},h)  \right]&\\
+\frac{\partial^2}{\partial \Delta h_k^2}\left[D^{(2)} (\Delta h_k,h) p(\Delta h_k| \Delta h_{k'},h)   \right],&
\label{fp}
\end{aligned}
\end{equation}
where the dependent variable is the height increment $\Delta h$ and
the independent variables is the time-lag (time scale)
$\tau_k$ or, respectively, $k$\cite{Friedrich2011}.
The surface elevation $h$ itself comes in as a second
independent variable.
This ensemble of Fokker-Planck equations is defined through
the extraction of the corresponding family of drift and diffusion
functions, $D^{(1)}(\Delta h_k,\tau_k,h)$  and 
$D^{(2)}(\Delta h_k,\tau_k,h)$, for the set of scales
$k$ as\cite{Risken}
\begin{equation}
  \begin{aligned}
    D^{(n)}(\Delta h_k,\tau_k,h_t)=&\\
    \lim_{\delta_\tau \to 0}\frac{1}{n!\delta_\tau}\langle [
    \Delta h_{\tau_{k}}-\Delta h_{\tau'_k}]^n\rangle \mid_{\Delta h_{\tau_{k}}} ,&
    \label{km_coef}
  \end{aligned}
\end{equation}
where $\tau'_{k} =  \tau_{k} + \delta_{\tau}$.
The Fokker-Planck equations provide the general multi-point statistics
of the data, enabling to generate new surrogate data as well as to
predict next wave events \cite{Ali2016}. 
As shown in the following, this stochastic description makes also possible
to set wave states in the framework of non-equilibrium thermodynamics
and its fluctuations theorems.

It is known that,
for a Markov process the integral fluctuation theorem (IFT) should hold,
i.e.~the balance between fluctuations that produce or consume entropy
is given by \cite{seifert2005}   %,Jarzynski2011}
\begin{equation}
\langle e^{-\Delta S} \rangle=1,
\label{IFT}
\end{equation}
where $\Delta S$ are entropy fluctuations and
$\langle \cdots \rangle$  is the expectation value over many
trajectories.
%Note that Eq.~(\ref{IFT}) implies that on
%average $\langle \Delta S \rangle > 0$, in agreement with the second
%law of thermodynamics.
%%%%%%%%%%%%%%%%%%%%%%%%%%%%%%%%%%%%%%
\begin{figure}[t]
\begin{center}
\includegraphics[width=0.4\textwidth]{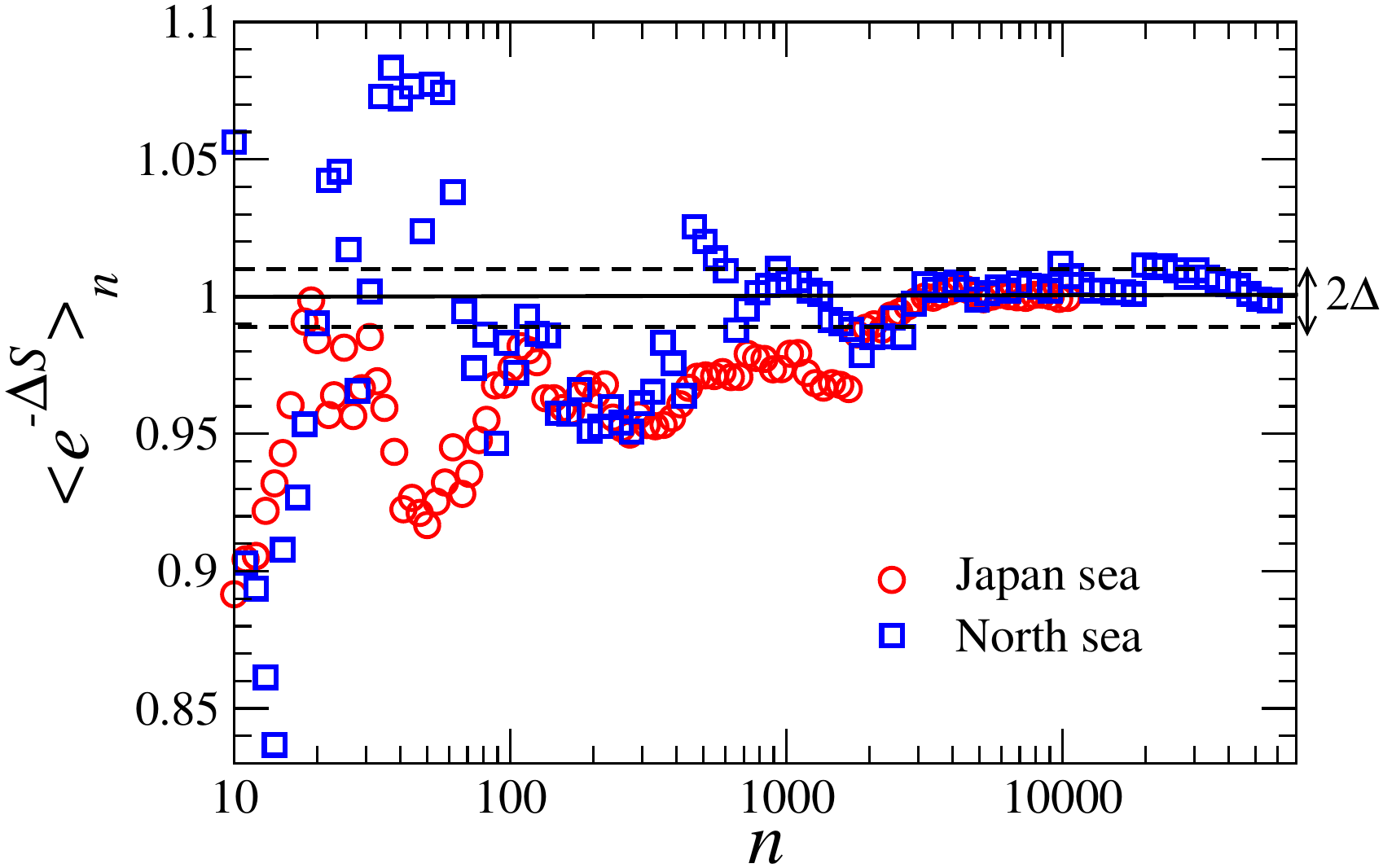}
\caption{\protect
  Rogue waves and extreme events of sea-level heights
  fulfill the integral fluctuation theorem (IFT),
  which means that statistically the entropy
  production fluctuates around zero.
  Here, the computation is applied to the data set from the
  Japan sea (circles) and the North sea (squares). For the
  former, since the modeling can be performed for arbitrarily
  large time windows, it is possible to observe the good
  convergence towards one ($\Delta\to 0$).
  In the case of these observational data sets one finds
  $\Delta\lesssim 1\%$.}
\label{fig02}
\end{center}
\end{figure}
%%%%%%%%%%%%%%%%%%%%%%%%%%%%%%%%%%%%%%
%%%%%%%%%%%%%%%%%%%%%%%%%%%%%%%%%%%%%%
\begin{figure}[t]
\begin{center}
\includegraphics[width=0.48\textwidth]{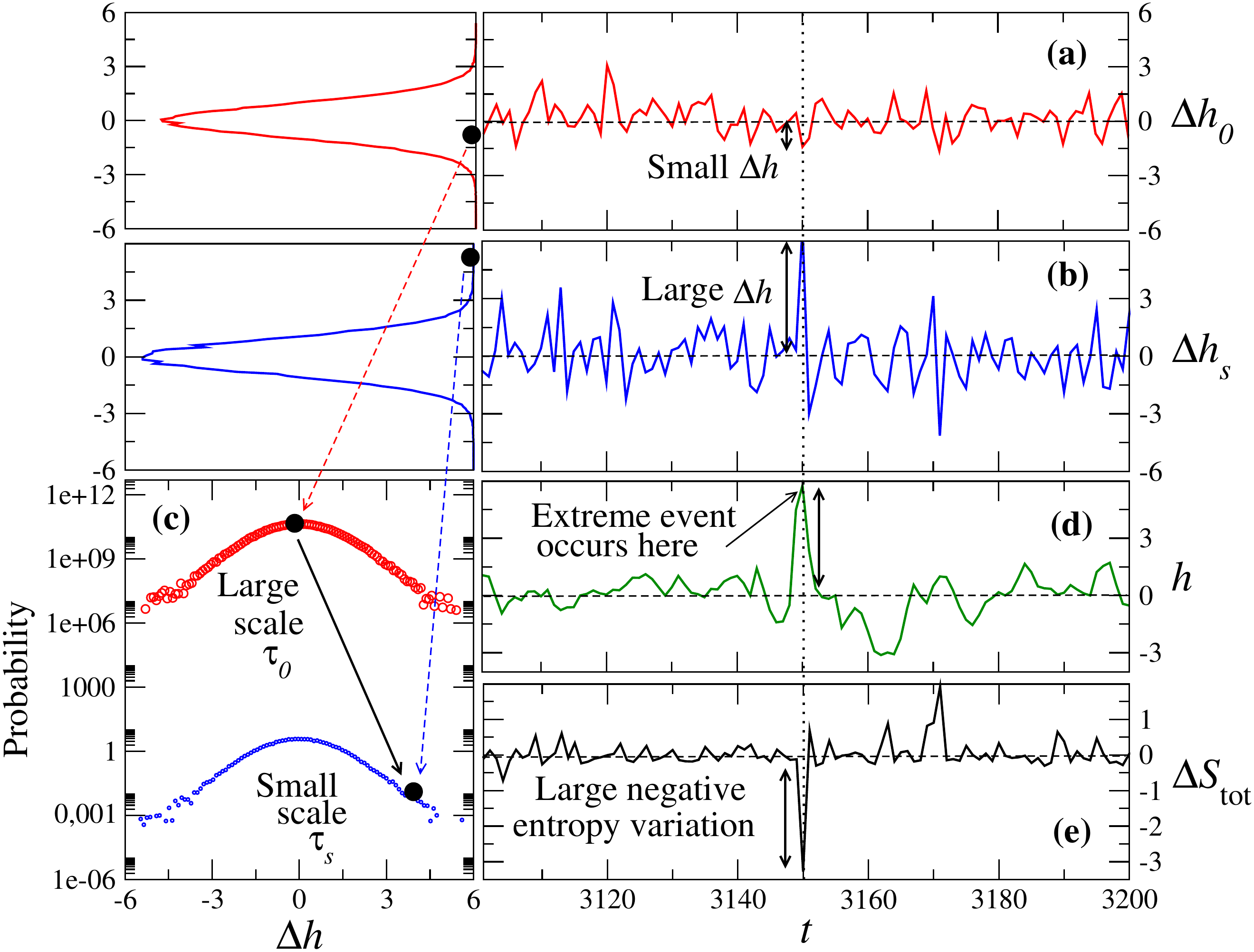}
\caption{\protect
         The close relation between an extreme event, i.e.~a large 
fluctuation of a given property, the wave height for rogue waves, within 
a short time interval, and the entropy production $\Delta S_{{\rm tot}}$:
the occurrence of an extreme event is identified by a {\it negative} 
entropy production, $\Delta S_{{\rm tot}}<0$.
         In {\bf (a-b)} we illustrate the time series of the
         height increments $\Delta h_{0,t}=h(t)-h(t-\tau_0)$
         (large scales)
         and $\Delta h_{s,t}=h(t)-h(t-\tau_s)$
         (small scales) respectively.
         On the left one sees the corresponding probability distribution
         for the increments, also plotted 
         (in logarithmic scale) in {\bf (c)} and shifted
         vertically for better comparison.
         {\bf (d)} The vertical dotted line marks the instant when
         a rogue wave event takes place: a large value of $h$ emerges.
         As one sees, {\bf (e)} the entropy variation
         is strongly negative, which indicates the statistical
         feature of a rogue wave or an extreme event in general.}
%         it has a small height increment at large time lags $\tau_0$ 
%         associated with large height increments at the smallest 
%         time lags $\tau_s$.}
\label{fig03}
\end{center}
\end{figure}
%%%%%%%%%%%%%%%%%%%%%%%%%%%%%%%%%%%%%%
%%%%%%%%%%%%%%%%%%%%%%%%%%%%%%%%%%%%%%
\begin{figure}[t]
\begin{center}
\includegraphics[width=0.45\textwidth]{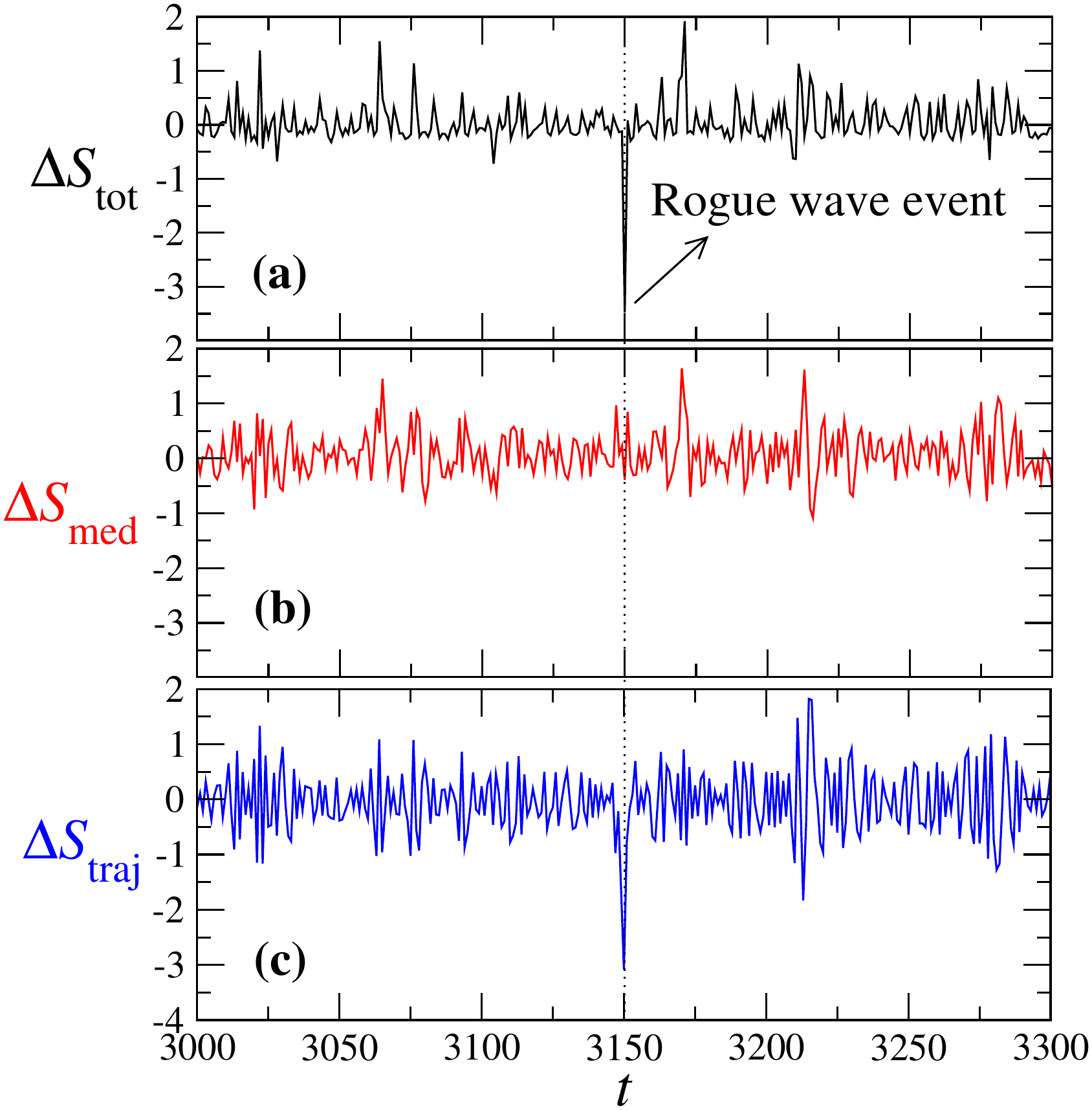}%
\caption{\protect
         {\bf (a)} With the total entropy production being negative whenever
         a rogue wave takes place, one sees that such negative
         variation of the entropy is not dominated by
         the variation of the entropy of the medium,  
         $\Delta S_{{\rm med}}$,  {\bf (b)},
         but rather by
         the variation of the individual trajectory's entropy only,
         $\Delta S_{{\rm traj}}$, {\bf (c)}.}
\label{fig04}
\end{center}
\end{figure}
%%%%%%%%%%%%%%%%%%%%%%%%%%%%%%%%%%%%%%%%%%%%%%%%%%%%%%%%%%%%%%%%%%%%%%%%%

Since the ocean wave system is Markovian in scale, the IFT should
also hold for increment ($\Delta h$)
trajectories in scale $k$. Based on the knowledge
of the Fokker-Planck equation for the system, a total entropy for each
increment trajectory can be defined\cite{seifert2005,nickelsen2013}.
This total entropy  is given by the sum of two contributions,
\begin{equation}
\Delta S_{{\rm tot}} = \Delta S_{{\rm med}} + \Delta S_{{\rm traj}},
\label{enttot}
\end{equation}
with $\Delta S_{{\rm med}}$ being the total entropy variation of
the surrounding environment,
which, between scales $k$ and $k-1$, is given 
by % \cite{nickelsen2013}
\begin{equation}
(\Delta S_{{\rm med}})_{k,k-1} = -\log{\left (
                         \frac{p(\Delta h_{k-1})}{p(\Delta h_k)}
                         \right )}.
\label{entsyst}
\end{equation}
The contribution
 $\Delta S_{{\rm traj}}$ is the entropy 
variation of the system, i.e.~along a specific trajectory
between those two time scales,
and is defined as \cite{seifert2005}
\begin{equation}
\begin{aligned}
  &(\Delta S_{{\rm traj}})_{k,k-1} =& \\
  &\int_{\tau_k}^{\tau_{k-1}} \frac{\partial}{\partial x} \Delta h(x)\frac{\partial}{\partial \Delta h}\log{\left ( p^{{\rm stat}}_{k} (\Delta h,x,h)\right )} dx ,&
\label{enttraj}
\end{aligned}
\end{equation}
where $p^{{\rm stat}}_{k}$ is the stationary solution 
of Eq.~(\ref{fp}) %the Fokker-Planck equation \cite{Friedrich2011}
\begin{equation}
  p^{{\rm stat}}_{k}(\Delta h,\tau,h)=\frac{1}{D^{(2)}(\Delta h,\tau,h)}\times
  \hbox{\large{e}}^{
    \int_{-\infty}^{\Delta h}\frac{D^{(1)}(x,\tau,h)}{D^{(2)}(x,\tau,h)} dx} .
  \label{statsol}
\end{equation}
To obtain $\Delta S_{{\rm tot}}$, the step-wise entropies
contributions, defined in Eqs.~(\ref{entsyst}) and (\ref{enttraj}),
have to be summed up along the scale trajectories.
Figure \ref{fig01}c shows the distribution of $\Delta S_{\rm tot}$
in each one of the data sets.

To show that IFT holds for both data sets we evaluate the
Eq.~(\ref{IFT}) for many events. As shown in Figure \ref{fig02} 
the IFT is fulfilled within an accuracy % of $\Delta \lesssim 1\%$ for
$\lesssim 1\%$ for 
 more than 2000 events.
The mean value $\langle e^{-\Delta S}\rangle$ changes very sensitively
with variations of the functional form of the Fokker-Planck equation.
Thus, the finding of the IFT can also be taken as a strong independent
support of the validity of our approach to characterize 
the complexity of wave states through scale processes.
In particular, it supports the thermostatistical description for the emergence of rogue waves
here proposed.

The convergence of the average 
$\langle e^{-\Delta S}\rangle$ is based on a sufficiently large data set,
as the exponential function puts much weight on rare negative entropy events.
This also means that due to the IFT there must be a special balance between 
events with positive and negative entropies. This leads us to the next point
to set the estimated entropy values for different increment trajectories in
connection with wave structures.

Figures \ref{fig03}a and b show time series of the height increments
at the largest and smallest scales, $\tau_0$ and $\tau_s$
respectively, with
the corresponding probability distributions (left plots).
In Fig.~\ref{fig03}c these probability densities are shown in a
semi-logarithmic presentation.
In Fig. \ref{fig03}d the part of the time series of the corresponding
wave height is show. 
The vertical dotted line marks the rogue wave seen in
Fig.~\ref{fig03}d, which is characterized by a small height
increment at the largest scale and a large increment
at the smallest scale. 
Figure \ref{fig03}c shows that the small increment at the
largest scale occurs with a high probability, while
the large increment at the smallest scale occurs with low
probability. 

To each wave height at a particular time instant $t$
belongs a scale trajectory  $\Delta h_{k,t}$ in $k$
($t$ is fixed).
Following such scale processes the total 
entropy variation $\Delta S_{{\rm tot}}$ can be positive 
(entropy production) or negative (entropy consumption).
Comparing the increment time series and the height 
time series with the series of the corresponding total entropy
variations (Fig.~\ref{fig03}e) one identifies an abrupt 
entropy consumption at the time of the occurrence of the
rogue wave.
%This is not mere coincidence: we found that rogue waves are always
%associated with large variations $\Delta h_s$ within short 
%time lags $\tau_s$ together with small variations $\Delta h_0$ 
%within the largest time scales $\tau_0$,
%and thus they result from an abrupt entropy consumption. 
As shown in Fig.~\ref{fig04},  it is not the entropy of
the environment but the entropy of the trajectory which becomes 
more negative and dominates the occurrence of the extreme event.
%The association of rogue waves with small $\Delta h_0$ and
%large $\Delta h_s$ is also quite intuitive, noting that rogue waves
%can be regarded as
%abrupt fluctuations strongly localized in short times. 
%******JoP - die Bedeutung des Satzes ist mit nicht klar******
% discussion of entropy distribution ********

Since the negative entropy variation shows to be an indicator of
an extreme event, one can now return to Fig.~\ref{fig01}c and use the
statistical distribution of total entropy variations for predicting
how reasonable it is to expect the occurrence of rogue waves
at a particular spot in the ocean.
The entropy values for the Japan Sea have a distribution shifted to
negative values, whereas the measurements taken for the North Sea
show a positive mode.
Assuming that the Laplacian distribution is a reasonable model
for $\Delta S_{\rm tot}$, we can now use the entropy value as a
measure for the likelihood of rogue waves.
To illustrate this fact, one can see 
from the distributions $p(\Delta S)$ in Fig.~\ref{fig01}c,
that 
an extreme event in the North sea
with an amplitude associated to an entropy
variation of e.g.~$\Delta S_{{\rm tot}} = -6$ is less likely to occur
than in the Japan sea by an factor of $10^{-4}$.

%Another interesting aspect, which can be deduced from the entropy distributions, is the detailed
%fluctuation theorem, or generalised Crooks fluctuation theorem stating that 
%$ p(-\Delta S) / p(\Delta S) = \exp{(-\Delta S)}$ , 
%which is a strong (sufficient) condition implying IFT\cite{seifert2012}. We find that both data set follow this exponential law, with different slopes in a semi-log presentation (see supplement part II). From this one may postulate that both wave states can be characterised by different temperatures, assuming 
%$\Delta S_{jap} / \Delta S_{north} = T_{north}/T_{jap}$.

In conclusion,
we introduce a thermodynamical approach, including the IFT, that
enables to interpret the emergence of rogue waves in the context
of the statistical physics and provides the possibility for estimate
the likelihood of them.
% simulation - coherent structures *****
An aspect that follows from our framework is that
the knowledge of the Fokker-Planck equation for the increment trajectories
given by the functions $D^{(1)}(\Delta h,\tau,h)$ and 
$D^{(2)}(\Delta h,\tau,h)$ can be used also for a simulation of
the sea surface elevation \cite{Hadjihosseini2014}
(see Supplementary Material I).
With such a model it is possible to generate
much longer time series to see which further patterns of rogue waves
may be expected.
Interestingly, preliminary simulations (see Supplementary Material II)
indicate that the patterns
obtained are qualitatively similar to those
obtained from deterministic modeling, like e.g. from
solving the non-linear Schr\"odinger equation, 
which is today considered a lowest order deterministic model for rogue waves 
in non-linear media \cite{Akhmediev2011a}.
These results strengthen the indication we provide above that
multi-point statistics for sea-level data includes typical features of
the alternative deterministic approaches, such as coherent structures.
The often debated difference between deterministic models and stochastic
approaches seem to fade out for the case investigated here, 
and can now serve as an inspiration
%, e.g.~for turbulent flows which
%as known are deterministically described by the Navier-Stokes equations
%but are commonly addressed from a statistical point of view.
%We expect our findings to find wide application
in all fields of science and technology where rare extreme events
are known to emerge from a complex dynamical system state, and where
at present statistical descriptions fall short of capturing key
properties and characterizing the emergence of the extreme events.
To name just a few fields where we think the approach presented will be
highly influential:
in fluid turbulence, where a hierarchy of simultaneous spatial and
time scales occurs, our approach might bridge the gap between the
statistical approaches to turbulent data and the %(deterministic)
Navier-Stokes
equations;
in solid mechanics, ranging from fracture processes,
acoustic emissions, up to earthquakes, the approach may be
applied, too;
and of course also for weather and climate extremes.
With our approach the study of rare and extreme events becomes
accessible for the wide toolbox of statistical physics, and a
completely novel way towards prediction has been opened.

\begin{acknowledgments}
  The authors thank A.~Chabchoub, A.~Engel, A.~Naert,
  %on the physics of rougue waves, as well as, 
%discussions on non-equilibrium thermodynamical interpretation with
%Andreas Engel, 
  D.~Nickelsen and N.~Reinke for useful discussions.
  %are acknowledged.
  This work was supported 
by VolkswagenStiftung (grant numbers 88480 and 88482) and Deutsche 
Forschungsgemeinschaft (3852/10).
The authors also thank the FINO1 Project, supported by the German Goverment through BMWi and PTJ, for providing the wind data 
measurements for the North Sea.
%the FINO1 Project is 
%\xxr{For helpful discussion we than A. Engel und A. Naert.}
\end{acknowledgments}

%%%%%%%%% THE BIBLIOGRAPHY
\bibliographystyle{apsrev4-1}
\bibliography{bibcite}

\newpage

\section*{Supplementary material I: Reconstruction of time series with extreme events}

Having a stochastic cascade model for the time series of surface heights in the ocean, with
or without rogue waves, we are now able to generate data that reproduces
all statistical features in the observations, including the ones related
to extreme waves.
%%%%%%%%%%%%%%%%%%%%%%%%%%%%%%%%%%%%%%
\begin{figure}[b]
\begin{center}
\includegraphics[width=0.4\textwidth]{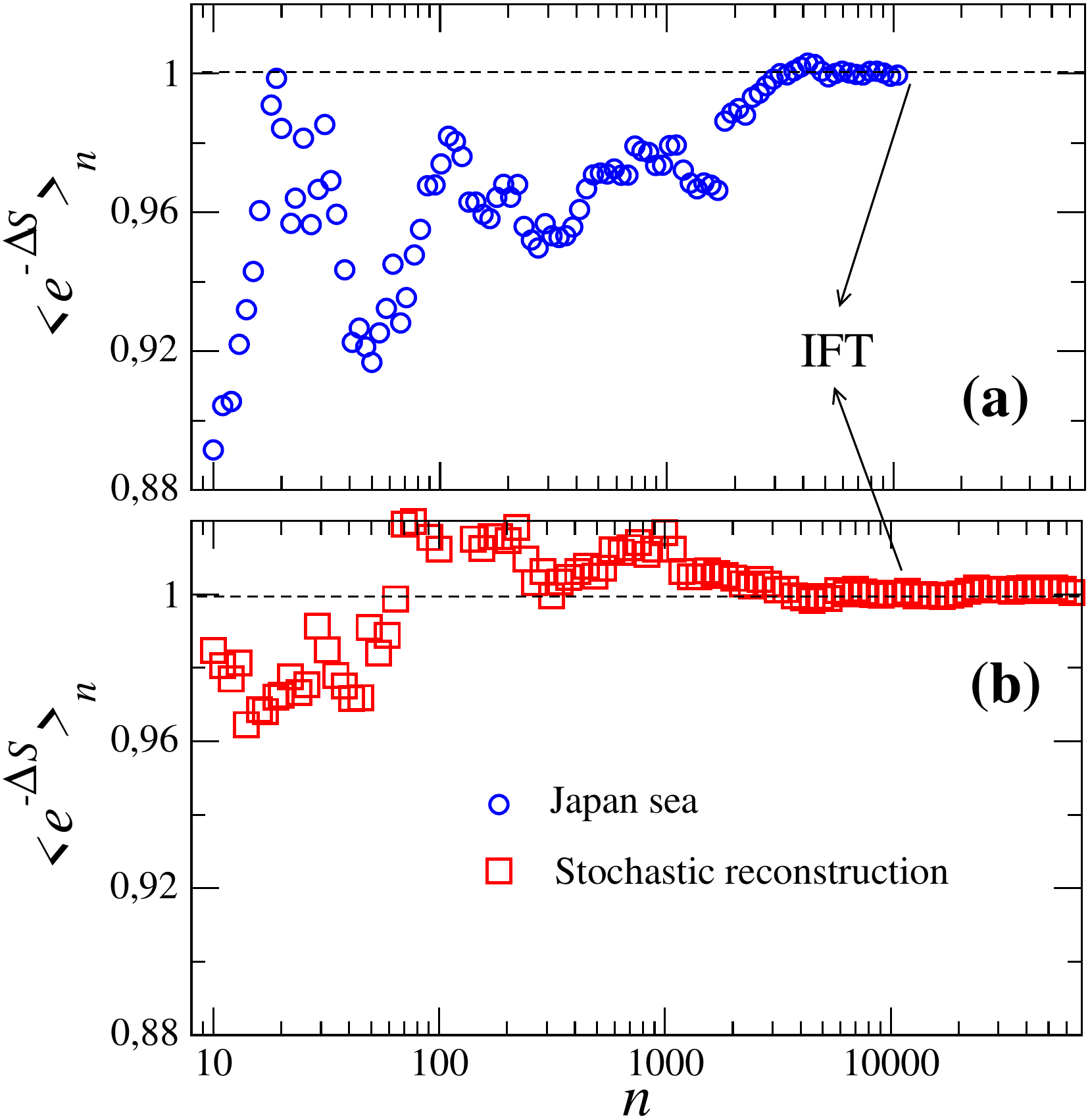}
\caption{\protect
         {\bf (a)} Rogue waves and extreme events in general fulfill
         the integral fluctuation theorem (IFT),
         which means that statistically the entropy
         production fluctuates around zero.
         {\bf (b)} The IFT also holds for the reconstructed data
         from the stochastic model of Japan Sea heights.
         Notice that, for the
         latter, since the modeling can be performed for arbitrarily
         large time windows, it is possible to observe the good
         convergence towards one.}
\label{fig05}
\end{center}
\end{figure}
%%%%%%%%%%%%%%%%%%%%%%%%%%%%%%%%%%%%%%

%The statistical reconstruction of
Such time series are generated with the $N$-point propagator as
mentioned in the article and shown in \cite{Ali2016}.
The starting point is that the process is Markovian in scale, for what
evidence is obtained by the verification of  $p(\Delta h_{k}\vert \Delta h_{k+1},\Delta h_{k+2},h)= p(\Delta h_{k}\vert \Delta h_{k+1},h)$. Assuming that the process is Markovian in scale the 
$(N+1)$-point propagator 
\begin{equation}
p(h_t\vert h_{t-\tau_1}, ... , h_{t-\tau_{N}}) =
\frac{p(h_t , h_{t-\tau_1}, ... , h_{t-\tau_{N}})}{p(h_{t-\tau_1}, ... , h_{t-\tau_{N}})} 
\label{cpdf2}%
\end{equation}
is used to construct the time series of heights. 
As the joint multi-point probabilities 
\begin{equation}
\begin{aligned} 
p(h_t , h_{t-\tau_1}, ... , h_{t-\tau_{N}})=
p(\Delta h_1, \Delta h_2,...\Delta h_N |h)\cdot p(h) 
\label{npoint}
\end{aligned}
\end{equation}
are equivalent to joint increment statistics, the
$(N+1)$-point propagator (of Eq.~(\ref{cpdf2})) can be expressed as
\begin{equation}
\begin{aligned}
&p(h_t\vert h_{t-\tau_1}, ... , h_{t-\tau_{N}})=&\\
  & \frac{\prod_{k=1}^{N-1} p(\Delta h_{k}|\Delta h_{k+1}, h_t )}
  {\prod_{k=2}^{N-1} p(\tilde{\Delta h}_{k},|\tilde{\Delta h}_{k+1},h_{t-\tau_1})}
  \times \frac{p(\Delta h_N|h_t)}{p(\tilde{\Delta h}_N|h_{t-\tau_1})}
  \times\frac{p(h_t)}{p(h_{t-\tau_1})} .&
\label{mp2}%
\end{aligned}
\end{equation} 
For reason of clarity we dropped the index $t$ for $\Delta h_{k}$, the two
increments $\Delta h_{k}$ and $\tilde{\Delta h}_k$ have two different
reference points, $h_t$ or, respectively $h_{t-\tau_1}$.
Here we used the Markov property to express all multi-point
distributions through simple
conditioned probabilities $p(\Delta h_{k}\vert \Delta h_{k'},h)$.
Note these conditioned increment probabilities are determined
by the Fokker-Planck equation, Eq.~(\ref{fp}),
for any reference value $h$.
The drift and diffusion coefficients are determined from the experimental
data, for details see \cite{Ali2016}.
Note that the Fokker-Planck equation is  continuous in $\tau$, to obtain
results for a finite step size, mentioned above, it has to be iterated over
this step.

Figure \ref{fig05} shows that the IFT is also fulfilled for the numerically reconstructed time series, which we obtain from the estimated Fokker-Planck equations. The same is observed for the North Sea data (not shown).
This is again a strong indication that all aspects of our stochastic methods are correct as well as the claim that the correct N-point statistics can be recovered.

\newpage

\section*{Supplementary material II: Statistics and coherent structures in extreme events}

\begin{figure}[b]
\begin{center}
\includegraphics[width=0.45\textwidth]{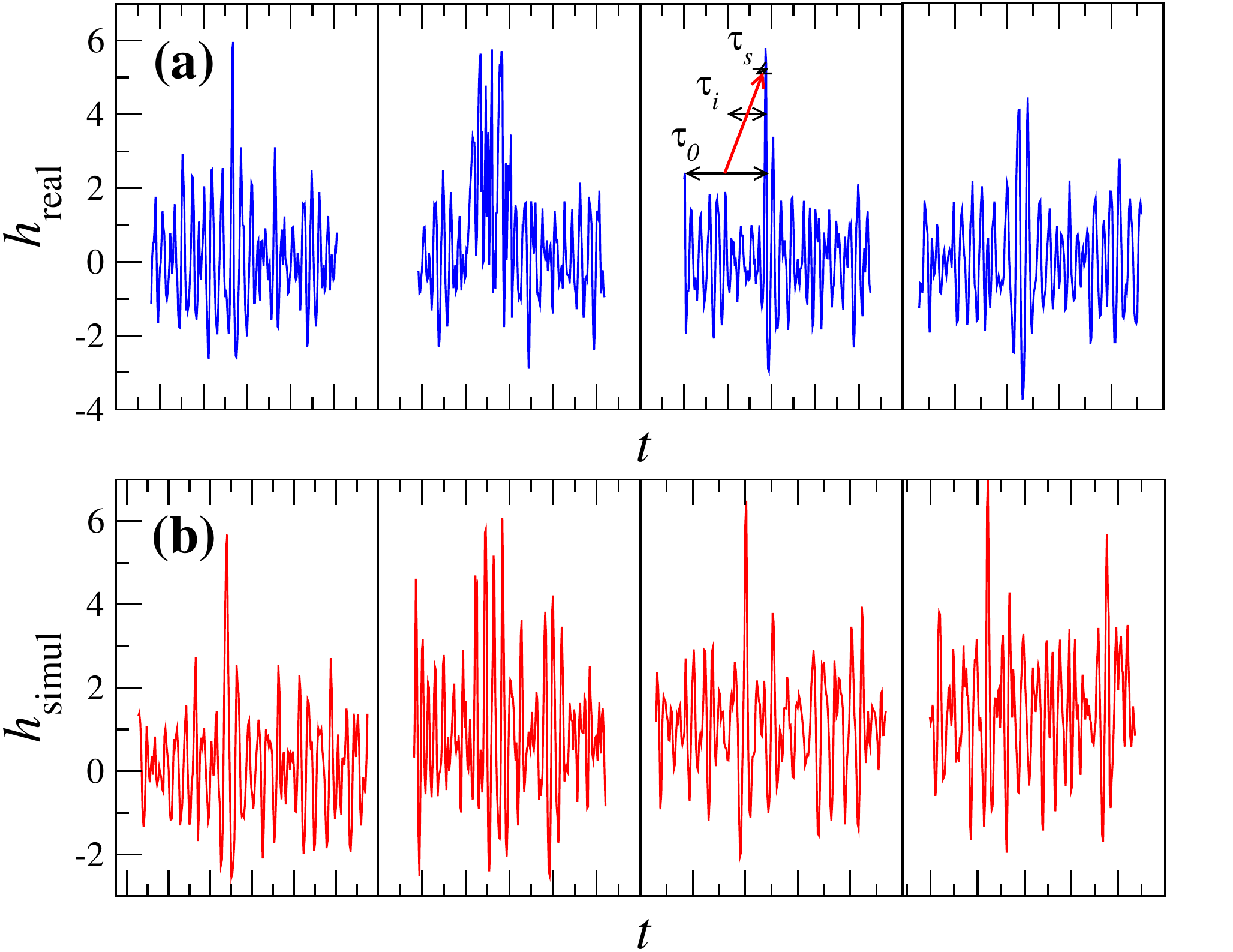}
\caption{\protect
         {\bf (a)} Illustration of pattern in sea-level height series 
         from the Japan Sea during which rogue waves are observed
         with {\bf (b)} the corresponding simulations using
         the stochastic framework based on a Fokker-Planck
         equation for the distribution of height increments.
         The similarity between simulated and measured patterns
         suggests an equivalence between the deterministic and
         the stochastic description based on what we call a
         {\it scale}-process
         of the height increments (see text).}
\label{fig06}
\end{center}
\end{figure}

As shown in Supplementary Material I,
the knowledge of the Fokker-Planck equation for the increment trajectories
given by  the functions $D^{(1)}(\Delta h,\tau,h)$ and 
$D^{(2)}(\Delta h,\tau,h)$ can be used for a simulation of
the sea surface elevation\cite{Hadjihosseini2014}.
It is possible to generate a time series much longer than the empirical
series to see which further patterns of rogue waves may be expected.
In sets of such simulated data one can actually observe events
that are very similar to real rogue waves. %already found elsewhere, as
As can be seen in Fig.~\ref{fig06}, the rogue wave patterns in our earlier
simulations, resemble the ones found in the Japan Sea\cite{Hadjihosseini2014}.
%%%%%%%%%%%%%%%%%%%%%%%%%%%%%%%%%%%%%%
%%%%%%%%%%%%%%%%%%%%%%%%%%%%%%%%%%%%%%

Interestingly, these patterns obtained are qualitatively similar
to those obtained from deterministic modeling, like e.g.~from
solving the non-linear Schr\"odinger equation, which is today
considered a lowest order deterministic model for rogue waves 
in non-linear media\cite{Akhmediev2011a}.
These results indicate that the stochastic approach of multi-point
statistics presented here also include typical features of the
deterministic description, like coherent structures.
Thus the often debated difference between deterministic models and
stochastic approaches seem to fade out at least for the case
investigated here.

\end{document}